\pdfoutput=1
\documentclass[a4paper, twoside, openany]{book}

\usepackage[svgnames]{xcolor} 
\usepackage{setspace}
\usepackage{pdfpages}
\usepackage[hidelinks]{hyperref}
\usepackage{xspace}
\usepackage[bottom=1in, top=1in, left=32mm, right=38mm]{geometry}

\usepackage{fancyhdr}
\usepackage{tocloft}

\makeatletter
\let\latexl@section\l@section
\def\l@section#1#2{\begingroup\let\numberline\@gobble\latexl@section{#1}{#2}\endgroup}

\let\latexl@chapter\l@chapter
\def\l@chapter#1#2{\begingroup\let\numberline\@gobble\latexl@chapter{#1}{#2}\endgroup}
\makeatother

\newcommand{\crule}[3][black]{\textcolor{#1}{\rule{#2}{#3}}}
\definecolor[named]{lipicsYellow}{rgb}{0.99,0.78,0.07}
\newcommand{\chbox}[0]{\crule[lipicsYellow]{0.65cm}{0.65cm}}

\begin{document}

\frontmatter

\pagestyle{empty}


\begin{titlepage} 

\begin{spacing}{2}

  \vspace*{3cm}
  {\huge
    
  \noindent
  \textsf{\textbf{21st International Workshop on}}

  \noindent
  \textsf{\textbf{Termination}}
  }
\end{spacing}

  \vspace*{1.5cm}
  \noindent
  {\large
    \textsf{\textbf{WST 2026, July 25, 2026}}}
  \par\noindent
  {\large\textsf{\textbf{affiliated with IJCAR at FLoC 2026}}}
  \par
  \vspace*{0.5cm}
  \noindent
  {\large
    \url{https://termination-portal.org/wiki/WST2026}
  }

\vspace*{3.0cm}

\begin{spacing}{2}
  \noindent
  \textsf{\large Edited by}~\\
  \textsf{\Large Florian Frohn and Étienne Payet}

\vspace*{9cm}
\noindent
\hfill\colorbox{lipicsYellow}{\Large \textsf{\textbf{\textcolor{black!70}{WST 2026~~~~~~~~~~~~~~~~~~~~~~~~~~Proceedings~~~~~~~~~~~~~~~~~~~~~~~~~~~~~~~~~~~~~~~~~}}}}

\end{spacing}
  \end{titlepage}

\cleardoublepage

\pagestyle{plain}
\pagenumbering{roman}
\setcounter{page}{1}



\chapter*{\chbox\xspace Preface}
\addcontentsline{toc}{chapter}{Preface}

This report contains the proceedings of the 21st International
Workshop on Termination (WST 2026), which was held in Lisbon on
July 25. It was affiliated with the 13th International Joint Conference on Automated Reasoning (IJCAR 2026), which was part of the Federated Logic Conference (FLoC 2026).

\medskip
The Workshop on Termination traditionally brings together, in an
informal setting, researchers interested in all aspects of
termination, whether this interest be practical or theoretical,
primary or derived. The workshop also provides a ground for
cross-fertilization of ideas from the different communities interested
in termination (e.g., working on computational mechanisms, programming
languages, software engineering, constraint solving, etc.). The
friendly atmosphere enables fruitful exchanges leading to joint
research and subsequent publications.

\medskip
The 21st International Workshop on Termination continues the
successful workshops held in St. Andrews (1993), La Bresse (1995), Ede
(1997), Dagstuhl (1999), Utrecht (2001), Valencia (2003), Aachen
(2004), Seattle (2006), Paris (2007), Leipzig (2009), Edinburgh
(2010), Obergurgl (2012), Bertinoro (2013), Vienna (2014), Obergurgl
(2016), Oxford (2018), the virtual space (2021), Haifa (2022), Obergurgl (2023), and Leipzig (2025).

\medskip
WST 2026 received 13 submissions.
After light reviewing the program committee decided to accept all
submissions.
Four submissions were presentation-only, i.e., the proceedings consist of nine papers.
The proceedings are published on arXiv.

\medskip

We would like to thank the program committee members
for their dedication and effort.

\vspace*{1cm}

\noindent
Florian Frohn and Étienne Payet



\chapter*{\chbox\xspace Organization}
\addcontentsline{toc}{chapter}{Organization}

\section*{Program Committee}

\bigskip
\begin{tabular}{l@{\qquad}l}
  Raúl Gutiérrez& Universitat Politécnica de Madrid\\
  Dieter Hofbauer & ASW Saarland\\
  Nils Lommen & RWTH Aachen University\\
  Johannes Niederhauser & University of Innsbruck\\
  Vincent van Oostrom & University of Sussex\\
  Hiroshi Unno & Tohoku University\\
  Wim Vanhoof & Université de Namur\\
  Florian Frohn (co-chair) & RWTH Aachen University\\
  Étienne Payet (co-chair) & Université de La Réunion
\end{tabular}

%

\bigskip


\newpage


\mainmatter
\pagestyle{fancy}
\pagenumbering{arabic}
\setcounter{page}{1}

\addtocontents{toc}{\vspace{0.5cm}}
\addtocontents{toc}{\textbf{\large Regular Papers}\par}

\newcommand\arXiv[3]{%
\addtocontents{toc}{\vspace{0.5cm}}
\addtocontents{toc}{{#2\\\emph{\mbox{}\hspace{1em} #3}}\protect\cftdotfill{\cftdotsep}\href{https://doi.org/10.48550/arXiv.#1}{arXiv:#1}\par}
}

\arXiv{2607.00521}
{Semantic Labelling in Practice}
{Dieter Hofbauer, Johannes Waldmann}

\arXiv{2604.20754}
{Termination of Innermost-Terminating Right-Linear Overlay Term Rewrite Systems}
{Naoki Nishida}

\arXiv{2605.29393}
{Unifying Semantic Path Order and Weighted Path Order}
{Teppei Saito, Nao Hirokawa}

\arXiv{2606.30127}
{Beyond Absolute Positiveness for Universally Quantified Non-Linear Polynomial Constraints}
{Carsten Fuhs}

\arXiv{2606.18977}
{PaSTTeL: Parallel analysiS framework for Termination and non-Termination of Lasso programs}
{Anissa Kheireddine, Souheib Baarir, Hugo De Sa Pereira Pinto}

\arXiv{2606.25448}
{Towards an HRS Category in TermCOMP}
{Johannes Niederhauser, Aart Middeldorp}

\arXiv{2606.23573}
{An Infinitary Lambda Calculus with Global Trace Condition}
{Stefano Berardi, Ugo de' Liguoro, Daisuke Kimura,\\Daniel Osorio-Valencia}

\arXiv{2606.17693}
{Verifying LTL for Infinite State Systems via Termination Analysis}
{Nils Lommen, Moritz Leven Rosarius, Jürgen Giesl}

\arXiv{2606.23516}
{Towards an Automated Reasoning Tool for Complexity Analysis of Automated Reasoners}
{Louis Rustenholz, Manuel V. Hermenegildo, Pedro Lopez-Garcia, Alessio Mansutti,\\Félix Ridoux, Niki Vazou}

\tableofcontents

\end{document}